\PassOptionsToPackage{unicode}{hyperref}
\PassOptionsToPackage{hyphens}{url}
\PassOptionsToPackage{dvipsnames,svgnames,x11names}{xcolor}
\documentclass[
  11pt,
]{article}
\usepackage{xcolor}
\usepackage[margin=1in]{geometry}
\usepackage{amsmath,amssymb}
\usepackage{iftex}
\ifPDFTeX
  \usepackage[T1]{fontenc}
  \usepackage[utf8]{inputenc}
  \usepackage{textcomp} 
\else 
  \usepackage{unicode-math} 
  \defaultfontfeatures{Scale=MatchLowercase}
  \defaultfontfeatures[\rmfamily]{Ligatures=TeX,Scale=1}
\fi
\usepackage{lmodern}
\ifPDFTeX\else
\fi
\IfFileExists{upquote.sty}{\usepackage{upquote}}{}
\IfFileExists{microtype.sty}{
  \usepackage[]{microtype}
  \UseMicrotypeSet[protrusion]{basicmath} 
}{}
\makeatletter
\@ifundefined{KOMAClassName}{
  \IfFileExists{parskip.sty}{%
    \usepackage{parskip}
  }{
    \setlength{\parindent}{0pt}
    \setlength{\parskip}{6pt plus 2pt minus 1pt}}
}{
  \KOMAoptions{parskip=half}}
\makeatother
\usepackage{longtable,booktabs,array}
\usepackage{caption}
\usepackage{calc} 
\usepackage{etoolbox}
\makeatletter
\patchcmd\longtable{\par}{\if@noskipsec\mbox{}\fi\par}{}{}
\makeatother
\providecommand{\tightlist}{%
  \setlength{\itemsep}{0pt}\setlength{\parskip}{0pt}}
\usepackage{graphicx}
\usepackage{float}
\pdfsuppressptexinfo=-1 \pdfinfoomitdate=1 \pdftrailerid{}
\makeatletter
\newcommand{\needspace}[1]{\par\penalty-100\begingroup\setlength{\dimen@}{#1}\dimen@ii\pagegoal\advance\dimen@ii-\pagetotal\ifdim\dimen@>\dimen@ii\ifdim\dimen@ii>\z@\vfil\fi\penalty-10000\fi\endgroup}
\makeatother

\newcommand{\FigFloat}[4]{%
  \begin{figure}[H]
  \centering\pdfliteral page{/Span <</Alt (#4)>> BDC}%
  \includegraphics[width=\textwidth]{figures/#1.pdf}%
  \pdfliteral page{EMC}\par
  \captionsetup{font=small}\caption{#2}\label{#3}
  \end{figure}}
\usepackage{bookmark}
\IfFileExists{xurl.sty}{\usepackage{xurl}}{} 
\makeatletter
\@ifundefined{xmpquote}{}{}
\makeatother
\hypersetup{
  pdftitle={Beyond Benchmarks: Can Small Open-Weight Models Get Real Coding Work Done?},
  pdfauthor={Vaibhav Jain, Independent Researcher, India},
  colorlinks=true,
  linkcolor={black},
  filecolor={Maroon},
  citecolor={Blue},
  urlcolor={blue},
  pdfcreator={LaTeX via pandoc}}

\title{Beyond Benchmarks: Can Small Open-Weight Models Get Real Coding
Work Done?}
\usepackage{etoolbox}
\makeatletter
\providecommand{\subtitle}[1]{
  \apptocmd{\@title}{\par {\large #1 \par}}{}{}
}
\makeatother
\subtitle{An Engineering Journey Through Model Serving, Tool Calling,
Runtime Fragmentation, and Verified Coding Tasks}
\author{Vaibhav Jain, Independent Researcher, India}
\date{}

\begin{document}
\maketitle

\subsection{Abstract}\label{abstract}

Making an open-weight language model generate text is not the same as
making it finish coding work. This retrospective empirical systems
experience report follows my engineering work from July to October 2026
across coding clients, model-specific serving recipes, vLLM, SGLang,
llama.cpp, and MLX. Models behind a nominally compatible endpoint still
differed in tool-call formatting, turn control, context behavior, cache
reuse, and reliability. A served model, coherent text, executed tool
calls, and code that passes independent behavioral checks proved to be
distinct levels of progress.

The evidence includes 4,165 coding-agent gateway turns from my own
sessions, 623 debug jobs, and three separate delegation studies with
hidden behavioral checks and a fixed Claude Opus 5.5 parent. Saved
cold-prefix requests had a median time to first token of about 69.4 s,
versus 2.8 s for high-reuse requests (observational). In a 45-trial
study, direct coding passed 15/15 and delegation to a quantized
Qwen3-Coder 30B-A3B worker passed 12/15 and 14/15. On a second 15-task
suite under a fixed artifact contract, the completed replacement cohort
had 45/45 for direct coding and 12/45 for a Nemotron Cascade 2 worker
(11 worker substitutions). Sixteen sessions stopped by a subscription
limit before any model call were replaced; counting them as failures
gives 37/45 and 8/45. A Qwen3-Coder worker passed 0/45. In both studies
with a direct arm, direct coding had lower API list-price-equivalent
frontier cost per verified success. A post-hoc diagnostic found that 19
of 36 artifacts rejected only for renamed parameters passed the hidden
checks unchanged; they are not counted as successes.

The contribution is an evidence-qualified account of serving, agent
protocol, session continuity, artifact validation, and verification in
practice. The studies are not pooled and do not rank models or runtimes.

\textbf{Keywords:} open-weight language models; coding agents; software
engineering; inference systems; tool calling; KV caching; model serving;
empirical experience report

\subsection{1. Introduction: From a Running Model to Finished
Work}\label{introduction-from-a-running-model-to-finished-work}

An early question was broader than coding agents: could ordinary laptops
pool their memory and computation to serve models too large for any one
machine, without requiring a specialized wired interconnect? I imagined
teammates, even in different locations, sharing access to a more capable
model instead of each developer needing enough memory to host it
independently. I initially expected that reaching a larger model might
improve the quality of coding assistance. The subsequent work exposed a
second question: even if the distributed model could answer, could it
support a productive coding agent? It quickly became apparent that ``the
model is running'' is a very weak success criterion. A checkpoint can
load and answer questions yet fail when a client asks it to inspect a
file, call a tool, keep a long session alive, or preserve a repository
contract.

I first tested that distributed idea with llama.cpp across laptops.
Cloud serving began while I was still experimenting with distributed
inference on Macs. As sharded chat proved more usable than sharded
coding-agent sessions, I gradually made single-node serving the default
and concentrated on coding-client integration, a simpler gateway, local
Mac deployments, and eventually verified coding tasks. Each transition
followed an engineering constraint: restarts and weight transfers slowed
distributed iteration; model-specific serving recipes did not travel
cleanly across runtimes; a compatible API did not always preserve an
agent's tool semantics; and long sessions depended on cache state
outside the model. The challenge was to make those pieces work together,
not merely find a checkpoint that produced good text. Figure 1
summarizes the engineering layers that emerged during this journey.

\FigFloat{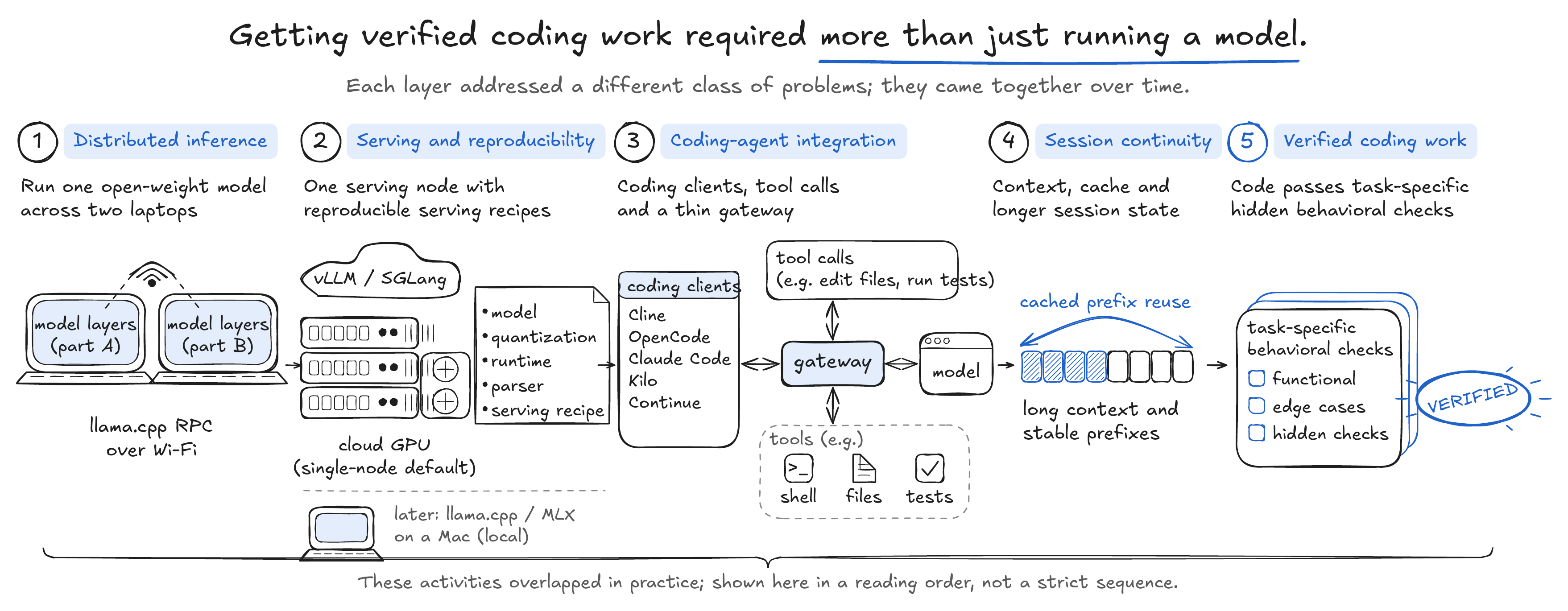}{Conceptual overview of the engineering layers that verified coding work required, July--October 2026: distributed inference across two laptops (Section 4), single-node serving with reproducible serving recipes (Section 5), coding-agent integration through a thin gateway (Section 6), session continuity (Section 7) and task-specific behavioral verification (Section 9). The layers are shown in a reading order, not a strict chronology; in practice they overlapped. Client names are those tried during the work; tool icons are examples.}{fig:1}{Five numbered columns from two laptops sharing a model, through a serving node and recipe, coding clients with a gateway and tools, a cached prefix, to task-specific checks stamped verified.}

For this report, a model's apparent progress can be separated into four
levels: \textbf{(1) served}, meaning the checkpoint loads and responds;
\textbf{(2) coherent text}, meaning it can answer in natural language;
\textbf{(3) parsed and executed tool calls}, meaning its responses fit
the agent's interface; and \textbf{(4) verified task completion},
meaning the resulting code passes task-specific behavioral checks. These
are not interchangeable. An endpoint can satisfy the first three while
failing the fourth; an inability to satisfy the third may prevent me
from measuring the fourth at all. Figure 2 shows these four levels and
the boundaries at which progress can stop.

\FigFloat{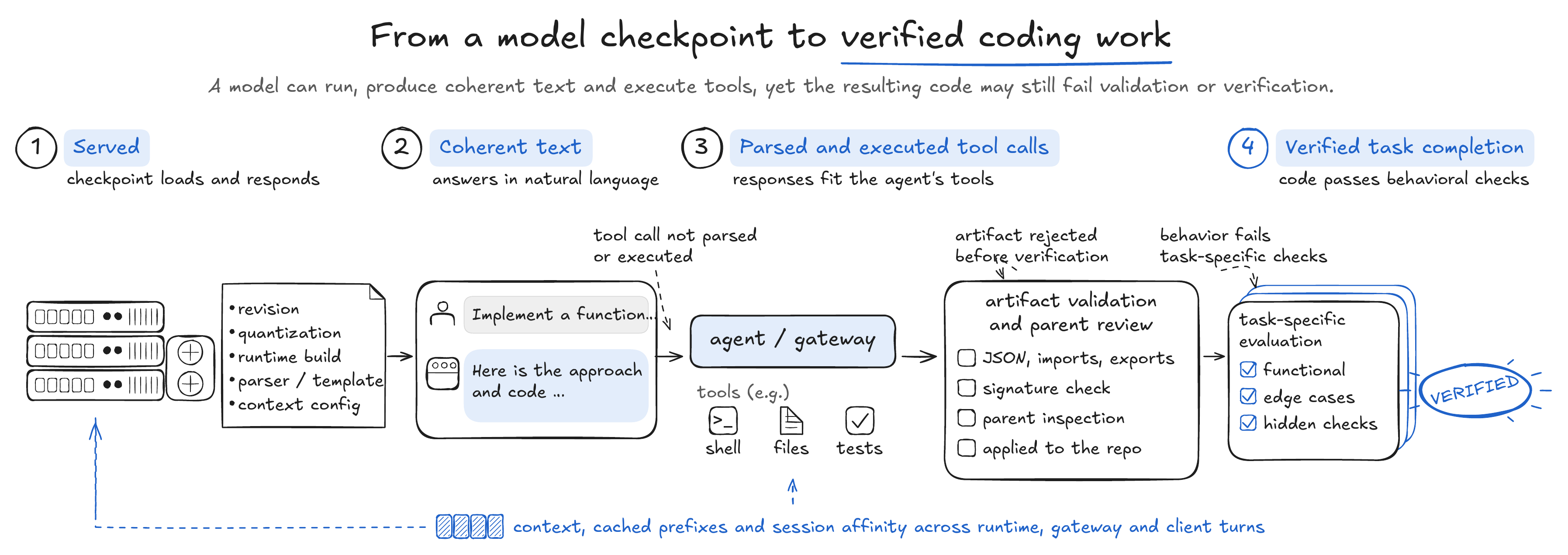}{The four success levels used in this report (Section 1) and where progress can stop. Artifact validation and parent review under the fixed artifact contract (Section 9.3) check form and interface (one JSON object, imports, exports and parameter signatures) before parent inspection and application; they are distinct from behavioral verification by task-specific hidden checks. The dashed band marks context, cached-prefix and session-affinity state, linked only to the serving runtime and the agent/gateway boundary (Section 7). Failure labels are engineering observations from Sections 5--9, not measured rates.}{fig:2}{Four numbered success levels left to right: a server with a serving recipe, a chat exchange, an agent gateway with tools, an artifact-validation checklist, and task-specific evaluation stamped verified, with a dashed cache band below.}

The later delegation studies added a further distinction. Working code
is not yet a verified outcome: in my delegation workflow, a worker's
artifact also had to satisfy an interface contract and be accepted and
applied by the parent before its hidden checks counted. A running model,
a functional implementation, an accepted artifact and a verified
end-to-end coding outcome are therefore different things. The clearest
illustration came from a post-hoc diagnostic run after the experiments
had closed (Section 9.4). I selected the 36 valid-JSON worker artifacts
that had been rejected solely for removing leading underscores from
parameter names. With only that rule bypassed, 19 of the 36 passed the
original hidden behavioral checks unchanged. They are not official
successes or verified substitutions, because they never went through
parent review and application. They do show that the design of an
interface contract can reject functionally correct code before normal
parent review begins. The diagnostic does not establish what any study
would have scored without that rule.

SWE-bench {[}1{]} and SWE-Lancer {[}2{]} motivate checking software
outcomes instead of isolated snippets. Work on harness design and
systems reliability {[}9, 11, 12{]} has already established the
importance of the surrounding infrastructure. I do not claim those
general principles as novel. My contribution is a concrete retrospective
account of runtime/recipe constraints, counted operational failures,
cache continuity, and bounded verified-task and delegation-economics
studies. These are case studies at the end of the engineering journey,
not a summary of the entire project.

The question throughout is: \textbf{what had to work between an
available open-weight checkpoint and a code change a developer could
trust?}

\subsection{2. Scope: What Counts as Useful
Work}\label{scope-what-counts-as-useful-work}

The report separates the \textbf{engineering work of making coding
agents operate} from the \textbf{bounded coding tasks evaluated with
hidden behavioral checks}. The former explains implementation choices
and operational failures; the latter provides directly scored outcomes
on selected TypeScript fixtures. Neither should be mistaken for a
representative evaluation of end-to-end software projects.

My target for local inference was a \textbf{working developer laptop},
not a machine cleared of other applications for a peak-throughput
benchmark. I wanted a model to run alongside Chrome, Cursor, Claude
Code, an editor, terminals and ordinary development processes. This
forced a recurring choice: use the most capable model I could fit, or
preserve more memory for context, KV cache, other applications and
responsive iteration. A larger quantized checkpoint could appear to fit
while leaving little room for a long coding-agent conversation; a
smaller one could restore headroom at the potential cost of task
capability. The relevant limit was not just the advertised maximum
context or weight-file size, but how much context and performance
remained practical under a real workload. This was a design objective
and an observed constraint, not a measurement of multitasking overhead.
In the later Nemotron and Qwen studies, the machine was a 32 GB Apple M1
Pro with substantial swap use, running one 4-bit 30B-class worker at a
time.

\textbf{Scope of ``small.''} In the title, \emph{small} is a
practitioner-facing shorthand for models substantially smaller in active
computation and deployment requirements than the frontier services used
as a reference. The primary tested worker here is a
\textbf{30B-total-parameter, approximately 3B-active-parameter
mixture-of-experts} Qwen3-Coder checkpoint, with later studies of
Nemotron Cascade 2 and Qwen3-Coder 30B-A3B, both 4-bit MLX checkpoints
of similar total and active scale. I do \textbf{not} define all models
below an arbitrary total-parameter threshold as equivalent, nor claim
results for dense 3B models, 70B models, or every open-weight checkpoint
I configured. The terms \emph{small} and \emph{smaller} in this paper
refer to these evaluated worker configurations unless explicitly noted
otherwise.

\subsection{3. Evidence, Workloads, and
Boundaries}\label{evidence-workloads-and-boundaries}

\subsubsection{3.1 Where the observations came
from}\label{where-the-observations-came-from}

This is a retrospective account of my own work, pieced together from
what I saved along the way: configurations and code changes,
integration-test results, inference logs, debugging conversations, and
the results of the coding study. I re-checked the reported counts
against those saved results rather than re-running experiments.
Throughout, I distinguish \textbf{operational measurements} (timings and
event counts), \textbf{task-verified outcomes} (results of hidden
behavioral checks), \textbf{implementation changes} (what I changed, not
necessarily what improved), and \textbf{engineering observations}
(problems I encountered without an independent measurement). These
records are private, so readers cannot check them directly; the Data
availability statement describes what is and is not shared.

These datasets have different units of observation and selection
procedures; the preserved summary does not establish that the Gemma
tool-request sample and debug-job sample are disjoint from, or exact
subsets of, the gateway-turn population.

The main sources of evidence were:

\begin{itemize}
\tightlist
\item
  \textbf{Structured coding evaluation (45-trial study):} 45 sessions
  across five TypeScript task categories and three execution conditions,
  scored using task-specific hidden behavioral tests separate from the
  coding agents' own success claims (described in Section 9).
\item
  \textbf{Second task suite:} 15 synthetic TypeScript repositories with
  sealed hidden checks. The Nemotron study paired 45 direct and 45
  delegated sessions; a separate Qwen follow-up ran 45 delegated
  sessions. Inputs, settings and model revisions were hash-recorded
  before scoring (Section 9.3).
\item
  \textbf{Additional task-level pilots:} payment implementation, bounded
  fee and invoice tasks, a twelve-adapter generation task, and earlier
  small-edit trials. Most individual cells have a single run and are
  analyzed as case studies rather than statistically stable rates.
\item
  \textbf{Operational turns:} 4,165 coding-agent turns logged by my
  gateway over three weeks in October 2026, spanning several model
  recipes. They include stop reasons, cache accounting, and TTFT on a
  subset of turns. These are my own real working sessions, not an
  experimental assignment of prompts to models.
\item
  \textbf{Debug jobs:} 623 unique jobs from selected saved debug
  sessions, predominantly Gemma 4 26B-A4B, including 85 narrowly defined
  cold-prefill records and 280 high-reuse records with performance
  metadata.
\item
  \textbf{Configuration and integration-test records:} runtime settings,
  prompt-prefix rules, the models I configured, and integration-test
  summaries for MLX, llama.cpp, vLLM, and SGLang.
\end{itemize}

I selected the models and workloads as practical problems arose, rather
than assigning identical tasks randomly across systems. Some saved
results also lack machine or runtime details. I therefore use those
records to explain engineering behavior, not to rank models by speed or
reliability. Section 12 discusses the resulting limits.

\subsubsection{3.2 What counts as a
result}\label{what-counts-as-a-result}

A coding trial is \textbf{verified} when its task-specific hidden
behavioral checks pass; these checks are separate from the coding
agent's own claims of success, not independently authored or externally
audited tests. An artifact being emitted, proposed, inspected, or
praised by the parent is not itself verification. A \textbf{successful
substitution} further requires the generated patch to be applied or
retained rather than replaced by frontier-authored code. A later check
of the saved outputs found that all 26 verified delegation trials kept
the worker-produced code. The harness alone does not prove that the
frontier parent made no intermediate file edits through shell commands;
therefore ``retained worker code'' must not be interpreted as ``no
parent edits of any kind.'' This distinction matters: some pilots passed
after the parent did not delegate, or without a recorded apply event. In
the later studies, a \textbf{verified substitution} additionally
required a proposed worker artifact, parent inspection, confirmed
application, a passing sealed check, and no subsequent parent source
rewrite. A proposed artifact rejected by the parent is not a
substitution.

Cost results use one measure throughout. \textbf{API
list-price-equivalent frontier cost} is an estimate calculated from
published API prices and the frontier model's recorded token usage. It
is not my actual subscription bill, and it excludes the unmeasured cost
of running local workers. After this definition I shorten it to
\textbf{frontier cost} where the meaning is unambiguous; Section 10
gives the price assumptions. A failed early stop can lower frontier cost
without yielding useful software, and is never counted as a saving.

\subsection{4. The Early Detour: Splitting Inference Across
Laptops}\label{the-early-detour-splitting-inference-across-laptops}

My original aim was not limited to splitting a 30B-class model between
two laptops. I wanted to know whether a larger collection of everyday
machines could jointly serve a model that none could hold alone,
potentially for colleagues on different networks. Avoiding special
cables or a shared physical setup was part of the hypothesis, not merely
a convenience: a solution that depended on specialized connections would
not support that distributed-team vision.

I began in late July with llama.cpp RPC, experimenting with
local-network connections, tunnels, a relay and WebRTC. Getting the
leader and helper to communicate on macOS was difficult: inbound
connections to the helper's RPC server failed or hung while outbound
connections worked. I made the helper dial out to the leader and used a
forwarding process with network access as another route; the precise
macOS security mechanism was never confirmed. One early configuration
also sent the helper's layers to its CPU instead of the Apple Silicon
Metal device. I corrected that device default, and later runs became
more responsive, although the layer split changed too, preventing
attribution to that correction alone. Once I had a working
configuration, I could use distributed inference over Wi-Fi for
conversational chat. That was a genuine feasibility milestone, but it
did not settle whether the system was responsive enough for a coding
agent.

The iteration loop was its own obstacle. In the RPC setup I used,
restarting after even a small configuration change sent model data from
the leader to the helper again: one saved 32B load transferred about 8.7
GB and took roughly five minutes over Wi-Fi. I tried retaining and
reattaching to live processes to avoid unnecessary reloads, and
experimented with many layer allocations rather than assuming that an
even division was best. Allocations derived from available memory
sometimes placed most of the model on the remote helper. An even-looking
layer allocation was not necessarily an even split of memory,
computation or execution time, and each experiment involved another slow
setup cycle. Some runs stalled or aborted; model transfers and
responsiveness varied with the connection. Cache reuse appeared
important during interactive use, but I did not isolate its effect from
networking.

This changed the question I was asking myself. \emph{If the model can
answer over Wi-Fi, is that enough for a developer agent that must make
repeated model turns around tool calls?} In my experience it was not.
Chat felt usable, while coding-agent sessions remained too slow and
cumbersome for productive work. In retrospect, a wired Thunderbolt link
might have been worth testing, but I did not measure it against Wi-Fi. I
also explored splitting mixture-of-experts computation at the expert
level, but did not establish a usable configuration. The preserved
record does not identify a specific upstream defect, so I cannot
attribute that outcome to one.

I eventually parked the multi-machine direction and moved to one serving
node at a time so I could make progress on the immediate coding-agent
problem. That decision reflected the performance and development-loop
costs of this particular implementation. It did not disprove distributed
inference or the broader idea of pooling machines, and because no
matched distributed-versus-single-node benchmark was preserved, it is
not a measured verdict on either.

\subsection{5. One Model, Several Runtimes: Deployment Was Not
Portable}\label{one-model-several-runtimes-deployment-was-not-portable}

Once I focused on single-node serving, the challenge shifted from
distributing a model to \textbf{making deployment repeatable}. I moved
through \textbf{vLLM} and \textbf{SGLang} on cloud GPUs, then returned
to \textbf{llama.cpp} with GGUF checkpoints on Apple Silicon in October
and added \textbf{MLX} later that month. The question was no longer just
which combination could load a checkpoint. Could I turn that combination
into a dependable deployment, recreate it after a machine change, and
still sustain a coding-agent session?

\textbf{Getting a model running was not simply a matter of downloading
weights and starting a server.} I encountered quantization and kernel
incompatibilities, mismatched FlashInfer and Python dependencies, GPU
driver problems, checkpoints larger than expected, context limits that
weren't passed through every layer, and runtime versions that needed
specific pins. On one cloud configuration, an absent compiler prevented
optimized attention kernels from building, so the runtime silently fell
back to other kernels until the compiler and backend configuration were
corrected; the surviving performance observation came from one
engineering note. On another machine shape, the wrong GPU driver could
lock up the host under CUDA load. A checkpoint that loaded under one
engine sometimes needed a different precision or revision under another.

These differences complicated the one-command setup scripts I was
building. One launch path crashed because its generated shell script
joined commands incorrectly; I fixed the joining and added a regression
test for that malformed form.

Worse, a shared package-installation change intended for one model
removed development headers another runtime required at boot. Existing
instances continued running, but freshly created machines failed until
the missing dependency was explicitly installed and a fresh-startup
regression test was added; an existing model-specific change rule was
also extended to cover shared setup steps.

The difficulties were therefore twofold: discovering a working
model-specific configuration and preserving it as a repeatable
installation rather than a one-off manual success.

On a laptop, selecting the model was another deployment problem rather
than a simple preference for the largest checkpoint. Quantization
reduced the space occupied by weights, but it did not eliminate memory
used by the runtime, KV cache or concurrent applications.
Mixture-of-experts models could have comparatively few active parameters
while still carrying substantial resident weights. Long contexts also
competed with free memory and prefill time. I repeatedly had to choose
between the potential ability of a larger model and the usable context,
responsiveness and headroom of a smaller one. I did not preserve a
controlled model-size-versus-context comparison, so these are trade-offs
I encountered, not a measured optimum.

The practical unit of deployment became \textbf{checkpoint revision $\times$
quantization $\times$ runtime build $\times$ parser/template $\times$ context configuration $\times$
agent client}. A successful recipe for one combination could not safely
be copied to another, and an initially successful startup was not yet an
operational deployment: the scripts also needed to reproduce the
environment and expose the behavior that a coding client expected.

\textbf{``Supported'' did not mean interchangeable.} Even after a model
loaded, I still had to align its parser, reasoning format and tool
schema with the coding client. That extra work is why I came to treat a
deployment recipe, and not a model name, as the unit of compatibility.

Compatibility at the API level only made a model reachable. Some models
narrated a command instead of calling a tool, described a planned action
as prose, or rewrote a function signature that should have remained
stable. I adjusted model recipes, prompts, delegation constraints and my
expectations rather than trying to rewrite their behavior inside the
gateway. These were practical adaptations, not tests of which prompt
worked better, and behavioral portability still had to be established
for each configuration.

\needspace{12\baselineskip}\textbf{Table 1: Selected model/runtime integration findings}

{\def\LTcaptype{none} 
\begin{longtable}[]{@{}
  >{\raggedright\arraybackslash}p{(\linewidth - 2\tabcolsep) * \real{0.5000}}
  >{\raggedright\arraybackslash}p{(\linewidth - 2\tabcolsep) * \real{0.5000}}@{}}
\toprule\noalign{}
\begin{minipage}[b]{\linewidth}\raggedright
Configuration
\end{minipage} & \begin{minipage}[b]{\linewidth}\raggedright
Integration finding
\end{minipage} \\
\midrule\noalign{}
\endhead
\bottomrule\noalign{}
\endlastfoot
GPT-OSS-120B / SGLang & Harmony tool parsing interacted with tool-choice
settings; output limits could truncate tool-call JSON (configuration
workaround; limited tests). \\
Laguna S 2.1 / SGLang & A checkpoint revision had to be pinned for the
documented 96-GB GPU configuration (configuration record; no comparative
experiment). \\
Laguna S 2.1 / vLLM & A missing compiler prevented FlashInfer JIT
kernels from building and caused fallback to other kernels (dated
configuration note; no retained trial series). \\
Qwen3.5-122B-A10B / SGLang & A thinking-budget processor serialization
problem blocked startup until the implementation was adjusted (recorded
fix; no verified coding score from this change). \\
Gemma 4 26B-A4B / MLX and llama.cpp & Template and parser anomalies
appeared in saved requests; later cache cohorts span both runtimes
(observational mixed-runtime records). \\
Qwen3-Coder 30B-A3B IQ3 / llama.cpp & Passed selected bounded coding
tasks; tested context settings differed from the model's native context
length (task-verified outcomes). \\
Nemotron Cascade 2 / MLX & Selected bounded tasks succeeded or failed;
some pilot worker records lack precise serving-artifact identification;
the later scored study recorded a fixed revision (bounded pilot
results). \\
\end{longtable}
}

These are selected integration cases, not a ranking of runtimes or
models; Sections 3 and 12 describe their evidence sources and limits.

During an October llama.cpp upgrade, I investigated tool-call parsing
and cache-reuse concerns, then rolled back the change the next day. Both
tool formatting and cache continuity mattered for the coding-agent
workflow, but the surviving record does not establish a specific
regression caused by that upgrade.

Dense and mixture-of-experts checkpoints further complicated deployment
because their usable precision, memory footprint, kernels and runtime
support differed. The practical question remained whether a specific
recipe could serve the model \emph{and} support the client behavior
required for coding.

\subsection{6. From Text Responses to Tool-Using Coding
Clients}\label{from-text-responses-to-tool-using-coding-clients}

The client side changed as well. I started with \textbf{Cline}, moved
within days to \textbf{OpenCode} (and tried Kilo and Continue), and from
September connected \textbf{Claude Code} through an Anthropic-compatible
gateway. In those early weeks I encountered malformed or text-only tool
calls, mismatched thinking channels, tool loops and missing session
identifiers. The symptoms differed enough that fixing one client or
model combination rarely solved every other combination.

At times, working with different checkpoints felt less like swapping
libraries and more like collaborating with people who had markedly
different habits. One model followed a tool contract literally; another
explained what it intended to do; another rewrote what it considered an
awkward interface. I do not mean that models are people or possess
personalities. The analogy captures an engineering surprise: unlike
exchanging one SDK implementation for another, using the same nominal
API with a different model did not guarantee the same behavior. That
made evaluation and adaptation part of integration itself.

That prompted a diagnostic question: \emph{when a coding agent fails to
use a tool, did the model fail, or did the surrounding system lose the
tool call on its way to execution?} In August I increasingly addressed
incompatibilities at their source (model templates, parser rules, serve
flags and request translation) instead of accumulating one-off output
repairs. In September I simplified the gateway toward a thin translator,
reducing the number of places where a request or response could be
altered. A later gateway audit made that principle concrete: validation
on the Claude path was restricted to protocol and security checks, while
model-quality interventions were removed. I also worked to propagate
upstream failures as errors rather than letting a failed stream appear
to finish normally with an empty response. None of this proved that
every failure was fixed; the actual client still had to be exercised end
to end.

To check the evolving translation layer before relying on the full
coding client, I exercised the gateway with a \textbf{Python API
harness} sending Anthropic Messages requests and Claude Code-style tools
to \textbf{Qwen3.5-122B-A10B on SGLang}. The harness was \emph{not} the
Claude Code application. Successive synthetic-workspace checks returned
different pass counts as the configurations evolved; the runs do not
isolate the effect of a single change. The preserved synthetic run
sequence was \textbf{2/12; a separate one-scenario 0/1; 10/12; 11/12;
12/12; and 12/12} passes. Separate existing-codebase checks returned
\textbf{1/3} and, in a later run with a different task mix,
\textbf{1/1}. I do not combine those results with the synthetic scores.
The existing-codebase material remains conditional on the
private-workload review in Section 12.

The real Claude Code sessions then exposed problems that passing an API
schema check had not eliminated. In one October episode, the client's
working directory failed to reach the model-facing request, leaving the
model to guess paths. In another, the gateway handled streamed
tool-argument fragments as replacements instead of appending them. Both
could derail an agent before the quality of its proposed code was even
tested. They are individual debugging episodes, not frequency estimates.

This changed how I thought about compatibility: I needed to test the
client, translation layer, model template and tool execution as one
path, including streamed arguments, stop behavior and session state.
Several concrete cases illustrate where that path broke:

\begin{itemize}
\tightlist
\item
  A GPT-OSS-120B SGLang path required coercing
  \texttt{tool\_choice=required} to \texttt{auto} because the former
  interacted badly with Harmony tool parsing. The serving configuration
  also used a higher output allowance on tool rounds to avoid truncating
  reasoning and tool JSON together. A recorded soak passed a tool suite
  while still encountering malformed arguments for a
  task-list-management tool (\texttt{todowrite}) and a plan-then-stop
  case.
\item
  Gemma 4 saved outputs included \textbf{eight} distinct malformed
  \texttt{\textless{}\textbar{}tool\_\allowbreak{}call\textgreater{}\allowbreak{}callcall:}-prefixed
  completions that produced no parsed tool call. The malformed
  completions illustrate a parser-boundary problem, not an overall
  agent-success rate. Three initially suspected leaked-token cases were
  excluded after a detection-rule recheck found no matching events.
\item
  Integration records distinguish recipes that are configured for a
  runtime from those actually served and evaluated. A model being
  configured is not evidence that it behaves correctly as an agent.
\end{itemize}

The gateway records made the distinction between \emph{tool-protocol
behavior} and \emph{task completion} particularly visible. The following
are event counts from selected saved requests, not pass/fail rates:

\needspace{12\baselineskip}\textbf{Table 2: Observed tool and gateway events (distinct
observational subsets)}

{\def\LTcaptype{none} 
\begin{longtable}[]{@{}
  >{\raggedright\arraybackslash}p{(\linewidth - 4\tabcolsep) * \real{0.3000}}
  >{\raggedright\arraybackslash}p{(\linewidth - 4\tabcolsep) * \real{0.3000}}
  >{\raggedleft\arraybackslash}p{(\linewidth - 4\tabcolsep) * \real{0.4000}}@{}}
\toprule\noalign{}
\begin{minipage}[b]{\linewidth}\raggedright
Source / denominator
\end{minipage} & \begin{minipage}[b]{\linewidth}\raggedright
Recorded outcome
\end{minipage} & \begin{minipage}[b]{\linewidth}\raggedleft
Count
\end{minipage} \\
\midrule\noalign{}
\endhead
\bottomrule\noalign{}
\endlastfoot
Gemma requests containing tools (492 prompts) & Parsed outputs ending in
\texttt{tool\_calls} & 294 \\
Gemma requests containing tools (492 prompts) & Outputs ending
\texttt{stop} with no parsed tool call & 144 \\
Gateway turns with recorded timing (2,816 turns) &
\texttt{empty\_content} & 74 \\
Gateway turns with recorded timing (2,816 turns) &
\texttt{stream\_failed} & 44 \\
Gateway turns with recorded timing (2,816 turns) &
\texttt{repetition\_cut} & 21 \\
Gateway turns with recorded timing (2,816 turns) & \texttt{job\_failed}
& 15 \\
Gateway turns with recorded timing (2,816 turns) & Fallback flag & 40 \\
\end{longtable}
}

The \textbf{294} tool-call outcomes and \textbf{144} no-tool-call
\texttt{stop} outcomes are selected recorded outcomes, not an exhaustive
partition of the \textbf{492} prompts; the remaining \textbf{54} have no
saved parsed response. A \texttt{stop} without a tool call can be an
appropriate final response. Each timed gateway turn records at most one
error class: the four listed error categories are disjoint (\textbf{154
of 2,816 turns}), with two rarer classes covering another \textbf{three}
turns. Fallback is a separate flag, and \textbf{19 of its 40} flagged
turns also had \texttt{empty\_content}. Without knowing what action each
task required, neither these events nor the stop reasons establish
coding success rates. What they \emph{do} show is that generation,
parsing and streaming can interrupt the agent before task correctness
can be evaluated. Prior controlled work has likewise documented harness
sensitivity and tool-interface effects {[}11, 12{]}.

\subsection{7. Long Sessions: Context and Cache State Are Part of the
Integration}\label{long-sessions-context-and-cache-state-are-part-of-the-integration}

Getting individual tools to work did not mean a long coding session
would remain usable. The question became: \emph{why could a conversation
that looked continuous to the developer suddenly behave as though the
model were starting over?} Two forms of state became important: the
client's evolving conversation and the runtime's cached token prefix.

A client or gateway that changes earlier messages, routes a session to a
different slot, or inserts auxiliary work can destroy cache continuity
without changing what the user thinks the conversation is about.

Long coding-agent conversations often reuse a large system prompt, tool
schema and the growing history of tool calls and results. An effective
inference backend can reuse cached key-value state for an
\emph{identical token prefix}, but application-layer prompt changes can
silently defeat that reuse. Practical changes in timestamps, system
banners, tools, or JSON serialization may force expensive prefill.

Compaction was a separate long-session challenge. In a private-workload
integration exercise, I deliberately lowered compaction thresholds in
\textbf{12 OpenCode/Qwen3.5-122B-A10B tests}. Sub-agent compaction
triggered in six (main-session compaction never triggered), but none of
those six sub-agent tasks completed; one recorded \texttt{PASS} label
even contradicted the incomplete-task checks. That showed why I could
not treat \emph{compaction occurred} as \emph{work completed}. This
example remains subject to the private-workload review in Section 12.

\subsubsection{7.1 Gemma session
observations}\label{gemma-session-observations}

\needspace{12\baselineskip}\textbf{Table 3: Gemma 4 26B-A4B requests from saved debug sessions (MLX
and llama.cpp)}

{\def\LTcaptype{none} 
\begin{longtable}[]{@{}
  >{\raggedright\arraybackslash}p{(\linewidth - 4\tabcolsep) * \real{0.2727}}
  >{\raggedleft\arraybackslash}p{(\linewidth - 4\tabcolsep) * \real{0.3636}}
  >{\raggedleft\arraybackslash}p{(\linewidth - 4\tabcolsep) * \real{0.3636}}@{}}
\toprule\noalign{}
\begin{minipage}[b]{\linewidth}\raggedright
Group definition
\end{minipage} & \begin{minipage}[b]{\linewidth}\raggedleft
Unique jobs
\end{minipage} & \begin{minipage}[b]{\linewidth}\raggedleft
Median TTFT
\end{minipage} \\
\midrule\noalign{}
\endhead
\bottomrule\noalign{}
\endlastfoot
Cold: no reused prefix (\texttt{cache\_n=0}), 30k--36k input tokens & 85
& 69.4 s \\
Warm: $\geq$90\% reported reuse, \textless2k fresh tokens & 280 & 2.8 s \\
\end{longtable}
}

These are selected, differently composed jobs rather than randomized or
matched runs; the timings do not isolate the effect of caching.

Among the 85 cold jobs, 54 fell between 65 and 82 seconds; 59 of the
cold jobs are attributed to llama.cpp. Of 280 warm jobs, 167 are
attributed to llama.cpp. The remaining 26 cold and 113 warm jobs are
attributed to MLX in the saved runtime breakdown. In warm jobs that
recorded decode throughput, the median was 38.5 tokens/s across 206
records. Some selected warm requests had TTFT near one second.

\subsubsection{7.2 Broader gateway-turn
observations}\label{broader-gateway-turn-observations}

\needspace{12\baselineskip}\textbf{Table 4: Selected gateway turns, not controlled model
comparisons}

{\def\LTcaptype{none} 
\begin{longtable}[]{@{}lll@{}}
\toprule\noalign{}
Recipe & Warm group (n), median TTFT & Cold group (n), median TTFT \\
\midrule\noalign{}
\endhead
\bottomrule\noalign{}
\endlastfoot
Gemma 4 26B-A4B & 1,128; 3.0 s & 259; 67.0 s \\
Nemotron Cascade 2 & 338; 10.8 s & 90; 86.6 s \\
Qwen3-Coder 30B-A3B IQ3 & 282; 2.0 s & 29; 135.6 s \\
\end{longtable}
}

In Table 4, ``warm'' means at least 8,000 reused prefix tokens
(\texttt{cache\_n\ \textgreater{}=\ 8000}), and ``cold'' means none
(\texttt{cache\_n\ ==\ 0}). \textbf{These definitions differ from Table
3}, where the warm group requires at least 90\% reported reuse and fewer
than 2,000 fresh tokens, while the cold group has no reused prefix and
30,000--36,000 input tokens. The model recipes were used under different
request mixes, and machine provenance is incomplete. The numbers show a
substantial \emph{observed} difference associated with prefix-cache
state; they do not rank model speed or show that any local patch caused
the difference.

\subsubsection{7.3 Application-layer cache
discipline}\label{application-layer-cache-discipline}

For sessions intended to reuse an exact cached token prefix, I treated
the model-facing message sequence as append-only where practical, except
when genuine context changes required rewriting earlier content. Earlier
tool messages should not be cosmetically re-serialized; streamed and
final tool call JSON must match; per-turn metadata should not modify the
frozen system prefix; and MLX cache fingerprints should include each
rendering input that genuinely affects output tokens while excluding
irrelevant changing fields.

Restoring hybrid sliding-window attention state from llama.cpp slot
checkpoints also proved awkward, so I added local handling for
restoration. One recorded restore-related slow case took approximately
\textbf{70 seconds}. I have no matched post-change timing; the immediate
engineering goal was to get the interrupted session back into a usable
state.

\subsubsection{7.4 Auxiliary requests and session
affinity}\label{auxiliary-requests-and-session-affinity}

Coding-agent traffic includes auxiliary requests, such as title
generation or permission-related operations, in addition to the main
coding conversation. In an inference setup with a single reusable
execution slot, an auxiliary request may replace the main conversation's
resident prompt state. The subsequent coding turn can then require
renewed prefill, even if the main conversation itself maintained a
stable prefix. In the saved debug sessions, 65 requests were classified
as auxiliary; 44 of these recorded no cached-prefix reuse. Among the
available high-reuse timing records, auxiliary requests had a median
TTFT of about 12.2 seconds (13 observations), compared with about 2.7
seconds for main requests (267 observations). These are differently
composed observational groups, \textbf{not} an estimate of the latency
caused by auxiliary requests or proof of a particular eviction event in
each case.

Session affinity was the other half of the problem. A stable prompt does
little good if the next turn reaches an execution slot without its
cached state. Because clients did not always supply a usable session
identifier, I added session-affinity and slot-handling logic. The
records do not isolate a speed gain from that change, but it led me to a
practical design principle: preserve both the token prefix and the route
to compatible cache state.

These findings complement established prefix/cache mechanisms such as
vLLM's PagedAttention {[}3{]} and SGLang's RadixAttention {[}4{]}. The
concern here is the \emph{application--runtime boundary}: even a
sophisticated cache cannot reuse a prefix that an upstream agent gateway
changes unnecessarily.

\subsection{8. When the Protocol Worked, Coding Could Still
Fail}\label{when-the-protocol-worked-coding-could-still-fail}

Once tool requests could pass through the gateway, a harder question
remained: \emph{what counts as completing the coding work: producing a
patch, type-checking it, or showing that it meets the requested
behavior?} I encountered a different class of failure: \textbf{the agent
could appear to make progress without completing the intended change}.
In October sessions, models sometimes described a shell command instead
of invoking it, narrated the tool they wanted to use, or reproduced a
long file path with characters missing. Those were observed
interactions, not measured rates.

\textbf{Repository understanding was another failure point.} In one
September Claude Code session, the agent re-implemented functionality
that already existed and left state that was never updated, even though
type checking did not flag the problem. This example survives only as a
contemporaneous account, not saved execution artifacts, and remains
subject to the Section 12 private-workload review. It illustrates why a
valid tool call and a successful type check need not imply the requested
behavior was implemented.

\textbf{Another failure class involved artifacts and repository
contracts.} In a bounded \textbf{invoice} attempt, a \textbf{Nemotron
Cascade 2} worker artifact failed export-preservation validation. The
delegation tool returned \texttt{malformed} before the parent inspected,
applied or tested any worker code; the saved status does not identify
which missing declaration triggered the check. That is a rejected
artifact, not a failed invoice behavioral test. In separate payment
pilots, outputs sometimes changed public function signatures (for
example, replacing \texttt{export\ function} declarations with
arrow-function constants), while the separate parent constraint list was
not forwarded to the worker in that period. Another Nemotron attempt
produced malformed JSON before any valid patch could be assessed.

\textbf{Execution limits could prevent evaluation altogether.} A
twelve-adapter \textbf{Nemotron Cascade 2} worker request stopped after
approximately \textbf{120 seconds} with an \texttt{unavailable/http}
result, produced no candidate implementation and left the stub
untouched. The direct version completed, but its estimated visible code
length fell below the task's intended \textbf{2,500-token} threshold for
testing a larger generated-code workload. This episode illustrates the
effect of timeout and workload boundaries on what can be evaluated, not
an established inability of the worker to implement the task.

\textbf{Finally, passing a scorer is not the same as attributing the
code to a worker.} In a payments pilot, a voluntarily delegated arm
passed despite \textbf{zero recorded delegation events}; another
scorer-passing pilot had no recorded inspect/apply events. Neither
establishes a verified worker substitution on its own. Across these
failure types, tracing the broken layer mattered as much as obtaining
candidate code, consistent with the broader systems view in {[}9{]}.

\subsection{9. Case Study: Coding Tasks with Hidden Behavioral
Verification}\label{case-study-coding-tasks-with-hidden-behavioral-verification}

I used Claude Code with Claude Opus 5.5 as the fixed parent harness
throughout these delegation studies. It was the client I had integrated
most deeply, and holding it fixed let me examine worker behavior and
delegation overhead within an actual workflow. OpenCode and other
harnesses might behave differently; I did not test them as
interchangeable parents, and these findings are about the tested system,
not harness-independent model capability.

The engineering journey eventually reached a more controlled question:
after the deployment, tool and session problems had been addressed well
enough to run bounded work, could an open-weight worker produce code
that passed task-specific checks? The later 45-trial evaluation is my
first repeated coding outcome dataset. I treat it as a \textbf{case
study} within the July-to-October integration work, not as its central
explanation.

\textbf{Study design.} The evaluation covered five TypeScript task
categories (\texttt{quote}, \texttt{prorate}, \texttt{customer},
\texttt{split}, \texttt{parts}), three conditions, three trials per
category and condition, totaling 45 sessions. The conditions were direct
frontier execution, optimized delegation, and batch-labeled adaptive
delegation. Task-specific hidden behavioral scorers, separate from the
agents' own success claims, determined verification. The harness uses a
Qwen3-Coder 30B-A3B IQ3 worker in the delegation conditions, with Claude
Opus 5.5 as the parent/frontier model. In this earlier study, the worker
received the artifact contract, task text and supplied file context, but
the parent's separate constraints field was not forwarded. A gateway
path also dropped a duplicate system message and truncated messages
above 1,800 characters. Of 41 reconstructed worker prompts, two lost
approximately 220 final characters; both belonged to failed trials. None
of the verified trials had a reconstructed first-attempt prompt over the
limit, although one retry prompt could only be estimated. These
observations leave the frozen outcomes unchanged but limit
interpretation of instruction-following failures. In the
\textbf{optimized} condition, the parent was instructed to delegate a
bounded implementation and apply worker artifacts rather than write
implementation code itself. In the \textbf{batch-labeled} condition, the
parent was likewise instructed to apply worker artifacts rather than
independently write the implementation; it could divide work into
sequential worker packages \textbf{when it judged task parts
independent}, or fall back to a single delegation for coupled changes.
Only \textbf{5/15} batch-labeled trials actually used a batch call
(three packages each); the remaining \textbf{10/15} used one delegation.
Among the 14 verified batch-labeled trials, five used batch calls and
nine used single delegation. Consequently, this condition is an
\textbf{adaptive prompting-and-delegation policy}, not a clean
experimental comparison of batching versus non-batching. These are the
study's own condition names, not standardized algorithms; worker
packages were not executed concurrently.

\subsubsection{9.1 Multi-trial comparison}\label{multi-trial-comparison}

\needspace{12\baselineskip}\textbf{Table 5: Structured coding-task verification (five task
categories, three trials per category per arm)}

{\def\LTcaptype{none} 
\begin{longtable}[]{@{}
  >{\raggedright\arraybackslash}p{(\linewidth - 8\tabcolsep) * \real{0.1765}}
  >{\raggedleft\arraybackslash}p{(\linewidth - 8\tabcolsep) * \real{0.2353}}
  >{\raggedleft\arraybackslash}p{(\linewidth - 8\tabcolsep) * \real{0.2353}}
  >{\raggedright\arraybackslash}p{(\linewidth - 8\tabcolsep) * \real{0.1765}}
  >{\raggedright\arraybackslash}p{(\linewidth - 8\tabcolsep) * \real{0.1765}}@{}}
\toprule\noalign{}
\begin{minipage}[b]{\linewidth}\raggedright
Execution condition
\end{minipage} & \begin{minipage}[b]{\linewidth}\raggedleft
Verified trials
\end{minipage} & \begin{minipage}[b]{\linewidth}\raggedleft
Total trials
\end{minipage} & \begin{minipage}[b]{\linewidth}\raggedright
95\% Wilson interval
\end{minipage} & \begin{minipage}[b]{\linewidth}\raggedright
Interpretation
\end{minipage} \\
\midrule\noalign{}
\endhead
\bottomrule\noalign{}
\endlastfoot
Direct frontier coding & 15 & 15 & {[}0.80, 1.00{]} & Every trial passed
the task-specific hidden behavioral scorer. \\
Optimized frontier $\rightarrow$ open-weight delegation & 12 & 15 & {[}0.55, 0.93{]}
& Three trials failed verification. \\
Batch-labeled adaptive delegation & 14 & 15 & {[}0.70, 0.99{]} & One
trial failed verification. \\
\end{longtable}
}

All \textbf{26} verified delegated trials retained worker-produced
implementations, although the harness does not rule out intermediate
parent edits. These are successes on the selected fixtures, not separate
applications. The Wilson intervals overlap; because the 15 trials per
condition repeat only five task categories, these run-level intervals
should not be interpreted as uncertainty across independent new coding
tasks. With three runs per task/condition cell and one worker
configuration, the data do not establish general success-rate
differences. The frontier-cost consequences are analyzed separately in
Section 10.

\subsubsection{9.2 Verified worker
contributions}\label{verified-worker-contributions}

Verified worker contributions were not limited to the later 45-trial
study. Earlier small-task experiments also produced retained worker
code, although they are less conclusive than the repeated evaluation.

A three-task comparison of two delegation arrangements reported 3/3
verified with retained worker code for the newer arrangement and 0/3 for
the earlier arrangement; the saved summary does not sufficiently
characterize their differences to isolate a mechanism. Separately, three
worker runs from the earlier small-edit trials passed hidden checks
\textbf{on the same three task types}; these are separate executions,
not six distinct task types. These single-run observations point to the
importance of orchestration, but cannot establish which architectural
change mattered. A later bounded \textbf{fee} task using
\textbf{Nemotron Cascade 2} also produced a task-verified open-weight
implementation that was applied to the task codebase. The fee example is
\textbf{one later, separate case study}, not the only successful worker
substitution across the project.

\subsubsection{9.3 A Second Task Suite Under a Fixed Artifact
Contract}\label{a-second-task-suite-under-a-fixed-artifact-contract}

I later prepared a second suite of 15 synthetic TypeScript repositories,
three each involving bug fixes, features, multi-file integrations,
export-preserving refactors and test-driven repairs, with sealed hidden
checks. The fixed artifact contract was derived from my practical
delegation setup; this was not a trial on production repositories. Each
original stub failed its check and each reference solution passed. These
tasks and checks were developed with AI assistance and hash-recorded
before the scored sessions.

The worker contract was fixed before scoring: one delegation attempt, a
120-second worker deadline, 4,096 output tokens, temperature 0, 32,768
model context and a 24,000-character supplied-file gate. The locally
served models used MLX on the 32 GB laptop. An artifact had to be one
JSON object, satisfy static checks on imports, exports and exported
function parameter lists, and pass parent inspection before application.
A passing sealed check without a worker application was not a verified
substitution.

The \textbf{Nemotron study} paired three direct and three Nemotron
Cascade 2 delegated repetitions for each task (90 sessions) on a frozen
schedule. The \textbf{Qwen follow-up} used the same tasks and fixed
artifact contract for 45 later delegated sessions with Qwen3-Coder
30B-A3B, but no new direct arm. These cohorts were not contemporaneous
or randomized together: the Qwen sessions ran later, and the parent's
prompt-cache state was not reset between studies. The earlier Qwen
IQ3/llama.cpp study used different tasks and should not be pooled with
them.

\needspace{12\baselineskip}\textbf{Table 5a: Verified outcomes on the second task suite}

{\def\LTcaptype{none} 
\begin{longtable}[]{@{}
  >{\raggedright\arraybackslash}p{(\linewidth - 6\tabcolsep) * \real{0.2143}}
  >{\raggedleft\arraybackslash}p{(\linewidth - 6\tabcolsep) * \real{0.2857}}
  >{\raggedleft\arraybackslash}p{(\linewidth - 6\tabcolsep) * \real{0.2857}}
  >{\raggedright\arraybackslash}p{(\linewidth - 6\tabcolsep) * \real{0.2143}}@{}}
\toprule\noalign{}
\begin{minipage}[b]{\linewidth}\raggedright
Study and condition
\end{minipage} & \begin{minipage}[b]{\linewidth}\raggedleft
Verified
\end{minipage} & \begin{minipage}[b]{\linewidth}\raggedleft
Verified worker substitutions
\end{minipage} & \begin{minipage}[b]{\linewidth}\raggedright
Task-mean verified rate (95\% Student-\emph{t} interval over 15 task
means)
\end{minipage} \\
\midrule\noalign{}
\endhead
\bottomrule\noalign{}
\endlastfoot
Nemotron study: direct Claude & 45/45 & n/a & 1.000 (degenerate) \\
Nemotron study: delegated Nemotron Cascade 2 & 12/45 & 11/45 & 0.267
(0.033--0.500) \\
Qwen follow-up: delegated Qwen3-Coder & 0/45 & 0/45 & 0.000
(degenerate) \\
\end{longtable}
}

\textbf{Task-level transparency.} The three repetitions on each
second-suite task were not independent new task types. Table 5b makes
the observed concentration visible; it reports verified task completions
(including Nemotron's one parent-rewrite success), not worker
substitutions. The direct arm passed all three repetitions of every
task. The Qwen follow-up passed none.

\needspace{12\baselineskip}\textbf{Table 5b: Verified sessions per second-suite task (out of three
repetitions)}

{\def\LTcaptype{none} 
\begin{longtable}[]{@{}
  >{\raggedright\arraybackslash}p{(\linewidth - 6\tabcolsep) * \real{0.2000}}
  >{\raggedleft\arraybackslash}p{(\linewidth - 6\tabcolsep) * \real{0.2667}}
  >{\raggedleft\arraybackslash}p{(\linewidth - 6\tabcolsep) * \real{0.2667}}
  >{\raggedleft\arraybackslash}p{(\linewidth - 6\tabcolsep) * \real{0.2667}}@{}}
\toprule\noalign{}
\begin{minipage}[b]{\linewidth}\raggedright
Second-suite task
\end{minipage} & \begin{minipage}[b]{\linewidth}\raggedleft
Direct verified / 3
\end{minipage} & \begin{minipage}[b]{\linewidth}\raggedleft
Nemotron verified / 3
\end{minipage} & \begin{minipage}[b]{\linewidth}\raggedleft
Qwen verified / 3
\end{minipage} \\
\midrule\noalign{}
\endhead
\bottomrule\noalign{}
\endlastfoot
\texttt{p3-01} & 3 & 3 & 0 \\
\texttt{p3-02} & 3 & 3 & 0 \\
\texttt{p3-03} & 3 & 1 & 0 \\
\texttt{p3-04} & 3 & 0 & 0 \\
\texttt{p3-05} & 3 & 0 & 0 \\
\texttt{p3-06} & 3 & 0 & 0 \\
\texttt{p3-07} & 3 & 0 & 0 \\
\texttt{p3-08} & 3 & 0 & 0 \\
\texttt{p3-09} & 3 & 0 & 0 \\
\texttt{p3-10} & 3 & 0 & 0 \\
\texttt{p3-11} & 3 & 3 & 0 \\
\texttt{p3-12} & 3 & 2 & 0 \\
\texttt{p3-13} & 3 & 0 & 0 \\
\texttt{p3-14} & 3 & 0 & 0 \\
\texttt{p3-15} & 3 & 0 & 0 \\
\end{longtable}
}

\needspace{12\baselineskip}\textbf{Table 5c: Outcomes of delegated sessions under the fixed
artifact contract}

{\def\LTcaptype{none} 
\begin{longtable}[]{@{}lrr@{}}
\toprule\noalign{}
Outcome & Nemotron & Qwen \\
\midrule\noalign{}
\endhead
\bottomrule\noalign{}
\endlastfoot
Verified worker substitution & 11 & 0 \\
Verified after parent source rewrite & 1 & 0 \\
Artifact rejected: parameter-signature validation & 13 & 26 \\
Artifact rejected: invalid JSON & 4 & 13 \\
Proposed artifact rejected by parent after inspection & 8 & 5 \\
Proposed artifact unreviewed after parent spend-limit stop & 1 & 0 \\
Worker deadline & 6 & 0 \\
Worker job cancelled & 1 & 1 \\
\textbf{Total} & \textbf{45} & \textbf{45} \\
\end{longtable}
}

Direct Claude verified all 45 sessions. Nemotron yielded 11 verified
substitutions, plus one session verified after the parent rewrote source
code; these are different outcomes. Qwen yielded no verified
substitutions. Crucially, 13 proposed artifacts (8 Nemotron, 5 Qwen)
were \textbf{rejected by the parent after inspection}, not applied and
then failed by the hidden tests. A separate Nemotron proposal went
unreviewed when the parent's subscription spend limit interrupted the
session. The sealed check in these cases ran on unchanged repositories.
Thus artifact production, parent acceptance and functional verification
must remain separate stages.

Because all 15 Qwen task means are zero, the computed Student-\emph{t}
interval is {[}0, 0{]}. This is a degenerate property of the observed
task means, not evidence that the underlying chance of success is zero.
Repetitions are also clustered within tasks and should not be treated as
45 independent task draws.

There was a procedural deviation in the Nemotron study: 16 scheduled
sessions rejected by a subscription spend limit \textbf{before any model
call} were excluded and replaced by fresh executions following a plan
change, contrary to the written protocol's instruction to retain every
scheduled session. An earlier session stopped \textbf{after} a worker
proposal and remained a failure rather than being rerun. The
replacements affected the reported outcomes: eight Nemotron reruns
contributed four of its 12 verified sessions, including three of its 11
substitutions; eight direct reruns contributed eight of the 45 direct
successes. If the 16 original pre-inference rejections were instead
counted as failures under the written keep-all-sessions rule, direct
coding would be 37/45, Nemotron verification 8/45 and Nemotron
substitutions 8/45. These alternative counts describe scheduled-workflow
interruptions, not additional model-inference failures. The principal
results describe the completed replacement cohort, with the deviation
and its effect disclosed. Figure 3 summarizes the verified outcomes and
frontier cost across the three separate studies.

\FigFloat{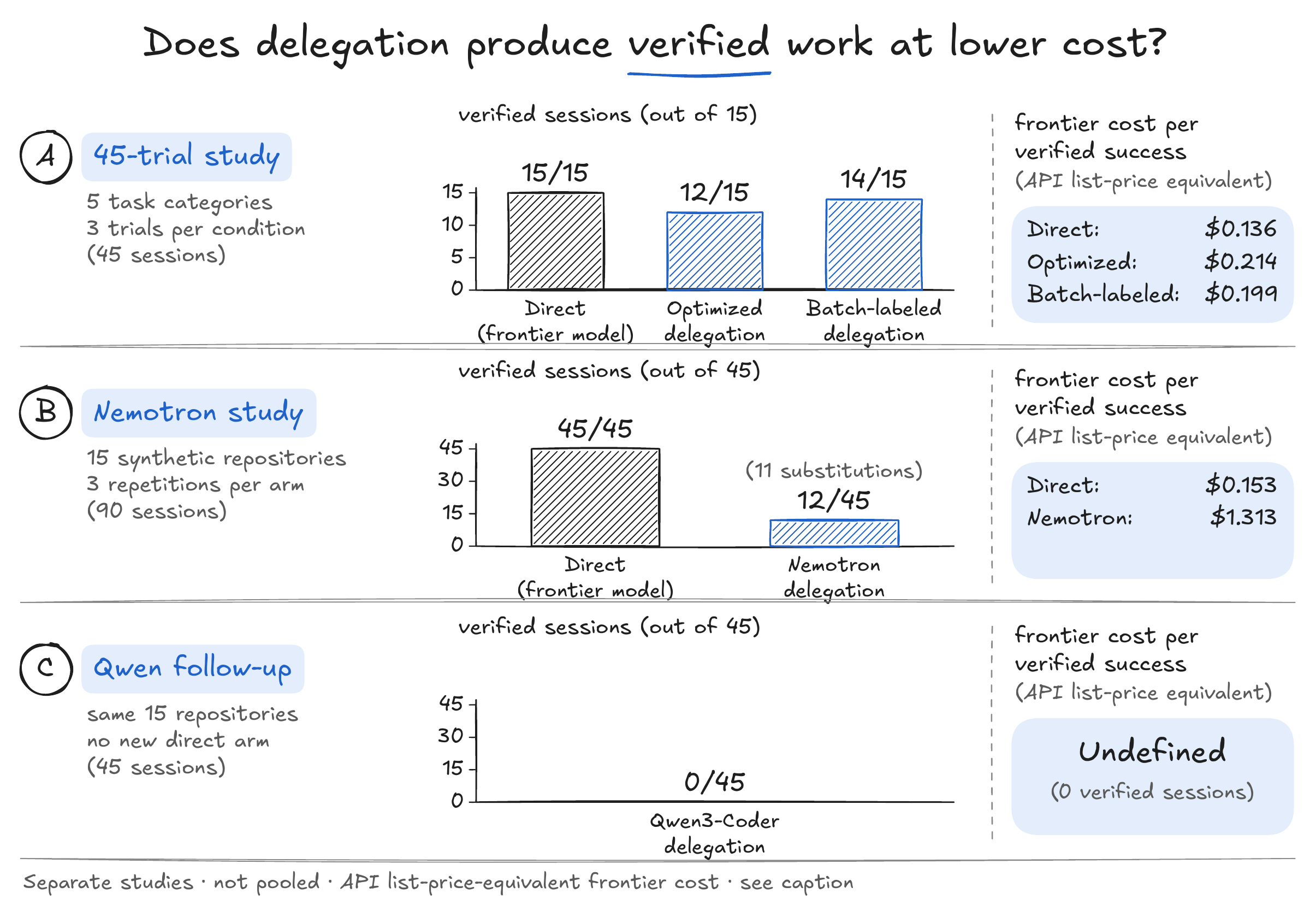}{Verified outcomes and API list-price-equivalent frontier cost in the three delegation studies, each with Claude Opus 5.5 in Claude Code as the parent. (A) The 45-trial study. (B) The Nemotron study and (C) the Qwen follow-up, on a separate 15-task suite under the fixed artifact contract. Panels use different tasks, workers, runtimes and dates and are not pooled. Panel (B) reports the completed replacement cohort; Nemotron's 12 verified sessions comprise 11 verified worker substitutions and one session verified after a parent source rewrite. The keep-all sensitivity (37/45 direct, 8/45 Nemotron) is given in Section 9.3. In (C) no session was verified, so cost per verified success is undefined. Costs include failed sessions; local worker cost was not measured.}{fig:3}{Three separate rows of verified-session bar charts with a frontier cost per verified success beside each; no values are pooled across rows.}

\subsubsection{9.4 When the Interface Rejects Correct
Code}\label{when-the-interface-rejects-correct-code}

The signature validator compared the text of existing exported function
parameter lists, including parameter names. Stubs often prefixed unused
parameters with an underscore (\texttt{\_entries}); workers sometimes
removed that underscore when implementing the function
(\texttt{entries}). Ordinary positional callers do not depend on this
spelling, but the contract's strict signature rule rejected the artifact
before parent review. This was compliance with the written contract, not
a demonstrated parsing bug.

After both studies were closed, I ran a \textbf{post-hoc offline
diagnostic}, without new model inference or code repair. I selected the
36 valid-JSON artifacts rejected solely for dropping leading underscores
(23 Qwen, 13 Nemotron), applied each unchanged to a fresh task copy
while bypassing only that name-comparison rule, and ran the original
sealed checks. Three other Qwen signature rejections changed more than
the underscores and were excluded. All 15 baseline tasks failed their
checks and all 15 reference implementations passed.

\needspace{12\baselineskip}\textbf{Table 5d: Post-hoc signature sensitivity (selected rejected
artifacts; not scored successes)}

{\def\LTcaptype{none} 
\begin{longtable}[]{@{}
  >{\raggedright\arraybackslash}p{(\linewidth - 6\tabcolsep) * \real{0.2000}}
  >{\raggedleft\arraybackslash}p{(\linewidth - 6\tabcolsep) * \real{0.2667}}
  >{\raggedleft\arraybackslash}p{(\linewidth - 6\tabcolsep) * \real{0.2667}}
  >{\raggedleft\arraybackslash}p{(\linewidth - 6\tabcolsep) * \real{0.2667}}@{}}
\toprule\noalign{}
\begin{minipage}[b]{\linewidth}\raggedright
Worker
\end{minipage} & \begin{minipage}[b]{\linewidth}\raggedleft
Underscore-only rejections
\end{minipage} & \begin{minipage}[b]{\linewidth}\raggedleft
Pass sealed check unchanged
\end{minipage} & \begin{minipage}[b]{\linewidth}\raggedleft
Still fail behaviorally
\end{minipage} \\
\midrule\noalign{}
\endhead
\bottomrule\noalign{}
\endlastfoot
Qwen3-Coder 30B-A3B & 23 & 12 & 11 \\
Nemotron Cascade 2 & 13 & 7 & 6 \\
\textbf{Total} & \textbf{36} & \textbf{19} & \textbf{17} \\
\end{longtable}
}

The diagnostic identified \textbf{19 functionally passing artifacts that
the original interface gate had rejected}, the observation previewed in
Section 1. On some tasks every artifact in this selected class passed;
on others, removing the gate exposed incorrect numeric results, thrown
errors, ordering mistakes or missed edge cases. The rule screened out
both useful and incorrect output.

These offline passes \textbf{are not verified substitutions or
counterfactual end-to-end successes}: the artifacts never entered normal
parent inspection and application. The frozen scores remain Qwen 0/45
and Nemotron 12/45. This narrower result demonstrates how a static
contract can suppress functionally passing code without establishing a
model-capability ranking or what the parent would have accepted. Figure
4 illustrates how interface-contract rejection differs from behavioral
failure, including the selected post-hoc diagnostic.

\FigFloat{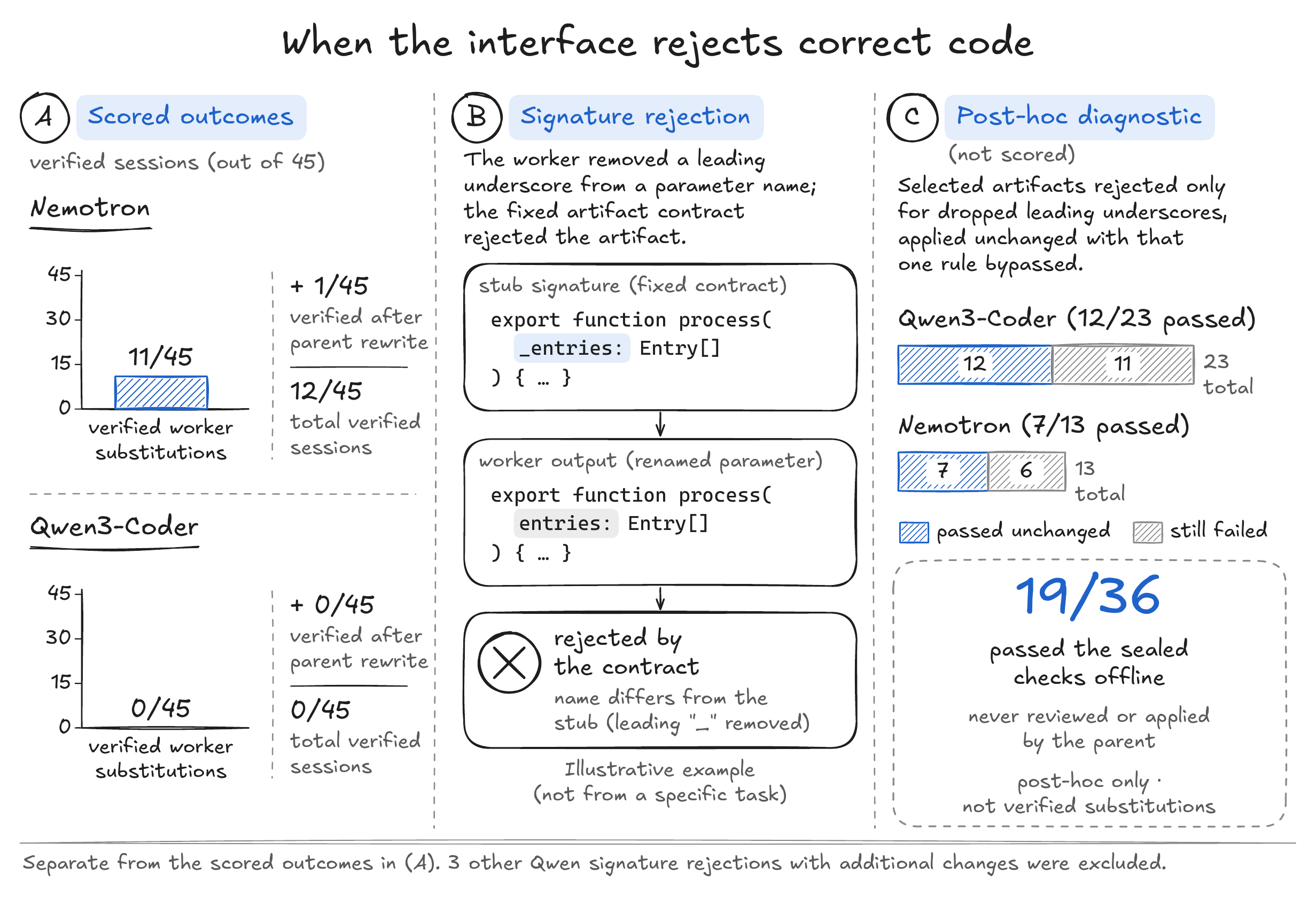}{Delegated-session outcomes under the fixed artifact contract and an offline diagnostic of one validation rule. (A) Scored outcomes (Table 5c): verified worker substitutions, sessions verified after a parent source rewrite, and their totals. (B) An illustrative example, not taken from a specific task, of the strict signature rule rejecting an artifact whose parameter name lost its leading underscore. (C) Post-hoc diagnostic (Section 9.4, Table 5d): the 36 valid-JSON artifacts rejected solely for dropped leading underscores were applied unchanged with only that rule bypassed, and 19 passed the same sealed hidden checks. These are not verified substitutions: the artifacts never entered parent inspection or application, and the scored results remain 12/45 (Nemotron) and 0/45 (Qwen). Three other Qwen signature rejections that changed more than the underscores were excluded.}{fig:4}{Three panels: scored verified counts per worker, an illustrative signature mismatch between a stub and a worker output, and a dashed post-hoc panel showing 19 of 36 selected artifacts passing offline.}

Among Qwen's 45 scored sessions, only 12 saved artifacts passed the
post-hoc behavioral checks unchanged after the underscore-only rule was
bypassed. Thus, relaxing only that rule, without repairing other
artifacts, could have made at most 12 of those existing artifacts
behaviorally passing candidates for further parent review. This is not a
projection of 12 verified substitutions: parent acceptance, application
and end-to-end verification were not replayed. Nor does it show that
Qwen's remaining sessions reflect fixed capability limits; other
interventions, such as the bounded repair attempt proposed in Section
11, were not tested.

\subsection{10. Case Study: The Cost of Delegating
Implementation}\label{case-study-the-cost-of-delegating-implementation}

The workflow tested a question that had become more concrete after the
integration work: \emph{if a smaller model can produce code that passes
verification, does delegating that work actually save anything?} A
strong parent model reads a repository and delegates a bounded
implementation to an inexpensive local worker; the parent inspects and
applies only suitable artifacts. Intuitively this could replace costly
frontier output tokens with local generation. But the parent must
communicate requirements, preserve constraints, inspect candidate
artifacts, and potentially recover from failures.

For a task, define:

\[C_{\mathrm{direct}} = C_{\mathrm{frontier,direct}}\]

\[C_{\mathrm{delegate}} = C_{\mathrm{parent,context}} + C_{\mathrm{spec}} + C_{\mathrm{review}} + C_{\mathrm{recovery}} + C_{\mathrm{worker}}\]

The equations describe a \textbf{conceptual full-cost comparison}. The
study measured the \emph{frontier} terms only; local worker expenditure
was unmeasured and must not be treated as zero. For a repeated workload,
compare \textbf{observed frontier cost per task-verified success} in
this study, not merely cost per request or cost for failed attempts that
ended early.

I priced the structured study using the study-era Claude Opus 5.5 list
rates: \textbf{\$4} per million uncached input tokens, \textbf{\$0.20}
per million cache-read tokens, \textbf{\$5} per million five-minute
cache-write tokens, \textbf{\$8} per million one-hour cache-write
tokens, and \textbf{\$20} per million output tokens. The accounting
distinguishes those token classes; thinking tokens are included in
output and are not charged twice. Each cache-write token was priced
using its recorded five-minute or one-hour cache lifetime (predominantly
one-hour). The resulting per-trial totals matched the client's reported
list-price cost estimates across all 45 trials. These are historical
study assumptions, not current vendor prices. All reported frontier
costs are \textbf{API list-price equivalents} based on recorded usage,
not per-token amounts paid: the frontier sessions ran under a personal
Claude subscription.

Across the five structured task categories, \textbf{direct execution had
the lowest measured frontier cost per verified success} (Table 6). The
amounts below include the frontier cost of failed trials in the
numerator; verification counts appear in parentheses.

\needspace{12\baselineskip}\textbf{Table 6: API list-price-equivalent frontier cost per verified
success by task category (USD)}

{\def\LTcaptype{none} 
\begin{longtable}[]{@{}
  >{\raggedright\arraybackslash}p{(\linewidth - 6\tabcolsep) * \real{0.2000}}
  >{\raggedleft\arraybackslash}p{(\linewidth - 6\tabcolsep) * \real{0.2667}}
  >{\raggedleft\arraybackslash}p{(\linewidth - 6\tabcolsep) * \real{0.2667}}
  >{\raggedleft\arraybackslash}p{(\linewidth - 6\tabcolsep) * \real{0.2667}}@{}}
\toprule\noalign{}
\begin{minipage}[b]{\linewidth}\raggedright
Task category
\end{minipage} & \begin{minipage}[b]{\linewidth}\raggedleft
Direct
\end{minipage} & \begin{minipage}[b]{\linewidth}\raggedleft
Optimized delegation
\end{minipage} & \begin{minipage}[b]{\linewidth}\raggedleft
Batch-labeled delegation
\end{minipage} \\
\midrule\noalign{}
\endhead
\bottomrule\noalign{}
\endlastfoot
\texttt{quote} & \$0.142 (3/3) & \$0.249 (2/3) & \$0.300 (2/3) \\
\texttt{prorate} & \$0.126 (3/3) & \$0.512 (1/3) & \$0.179 (3/3) \\
\texttt{customer} & \$0.156 (3/3) & \$0.204 (3/3) & \$0.210 (3/3) \\
\texttt{split} & \$0.117 (3/3) & \$0.153 (3/3) & \$0.157 (3/3) \\
\texttt{parts} & \$0.140 (3/3) & \$0.163 (3/3) & \$0.185 (3/3) \\
\textbf{All categories} & \textbf{\$0.136 (15/15)} & \textbf{\$0.214
(12/15)} & \textbf{\$0.199 (14/15)} \\
\end{longtable}
}

\emph{Each value is the total API list-price-equivalent frontier cost
for its trials, including failed trials, divided by the number verified;
parentheses show verified/total. Each task-category cell has only three
trials, and the optimized \texttt{prorate} figure rests on one verified
success. These are descriptive figures, not general performance
estimates.}

Direct execution also had the lowest mean frontier cost per trial in
each category, so the directional result does not depend on penalizing
failed attempts. The separate \textbf{fee} pilot illustrates the
distinction between successful substitution and savings: its worker
implementation passed and was applied, but the frontier cost was
\textbf{\$0.1856 with delegation} versus \textbf{\$0.1293 directly}, and
recorded wall time, which includes local worker execution, was
\textbf{126.4 seconds} versus \textbf{27.4 seconds}. These figures
describe that one case.

The second task suite strengthens this economic observation without
making it a general claim about model cascades.

\needspace{12\baselineskip}\textbf{Table 7: API list-price-equivalent frontier cost on the second
task suite (USD)}

{\def\LTcaptype{none} 
\begin{longtable}[]{@{}
  >{\raggedright\arraybackslash}p{(\linewidth - 8\tabcolsep) * \real{0.1579}}
  >{\raggedleft\arraybackslash}p{(\linewidth - 8\tabcolsep) * \real{0.2105}}
  >{\raggedleft\arraybackslash}p{(\linewidth - 8\tabcolsep) * \real{0.2105}}
  >{\raggedleft\arraybackslash}p{(\linewidth - 8\tabcolsep) * \real{0.2105}}
  >{\raggedleft\arraybackslash}p{(\linewidth - 8\tabcolsep) * \real{0.2105}}@{}}
\toprule\noalign{}
\begin{minipage}[b]{\linewidth}\raggedright
Condition
\end{minipage} & \begin{minipage}[b]{\linewidth}\raggedleft
Total API list-price equivalent
\end{minipage} & \begin{minipage}[b]{\linewidth}\raggedleft
Per session
\end{minipage} & \begin{minipage}[b]{\linewidth}\raggedleft
Per verified success
\end{minipage} & \begin{minipage}[b]{\linewidth}\raggedleft
Per verified substitution
\end{minipage} \\
\midrule\noalign{}
\endhead
\bottomrule\noalign{}
\endlastfoot
Nemotron study: direct & \$6.888 & \$0.153 & \$0.153 & n/a \\
Nemotron study: delegation & \$15.761 & \$0.350 & \$1.313 & \$1.433 \\
Qwen follow-up: delegation & \$16.023 & \$0.356 & Undefined (0 verified)
& Undefined \\
\end{longtable}
}

The Nemotron delegation arrangement consumed roughly \textbf{8.6 times}
the direct frontier cost per verified success. Failures are included in
the spending numerator. Qwen's per-success amount is undefined rather
than zero. Neither table measures local worker electricity, hardware,
depreciation or engineering time. Table 7 reports the completed
replacement cohort; the keep-all sensitivity in Section 9.3 gives
alternative outcome counts, not a separate cost calculation.

The result illustrates a practical economic tension: reducing frontier
implementation tokens does not guarantee lower frontier cost if
specifying, reviewing and recovering delegated work uses additional
parent tokens. The records cannot apportion those overheads cleanly into
separate orchestration and review effects. The finding applies to the
selected tasks, study-era parent pricing, worker configuration and
delegation contract.

This result does not refute cascades or model routing in general. Prior
work examines cost-aware model selection {[}5{]}, and Scrouting {[}13{]}
provides a positive counterexample: a small repository-scouting model
followed by routed frontier fixers solved 159 of 266 tasks versus 158
for the strongest single fixer, at approximately one fifth of its
reported cost per solve. Its
\textbf{small-model-scouts/frontier-model-fixes} direction is the
reverse of my \textbf{frontier-specifies/local-model-implements}
contract. I therefore limit the negative cost finding to the measured
delegation arrangement and do not generalize it to all routing or
cascades. An architecture with substantially different task boundaries
or a large mechanically generated artifact might behave differently, but
my data cannot establish that counterfactual.

\subsection{11. Practical Lessons Across the
Journey}\label{practical-lessons-across-the-journey}

The journey suggests five practical rules. They are not new universal
principles; they emerge from trying to make the components work in one
evolving setup and from the bounded studies reported here.

\textbf{First, define success at the task boundary.} Serving a
checkpoint, obtaining a coherent answer, executing parsed tools, and
producing verified code are distinct milestones. Test the actual client
and the resulting behavior, rather than trusting a plausible response or
a harness's own \texttt{PASS} label.

\textbf{Second, treat deployment as a full recipe and model selection as
a workload decision.} Checkpoint revision, quantization, runtime,
template, context, gateway, client and resident memory all matter. On a
working laptop I could choose a larger, potentially stronger model or
leave more room for KV cache, useful context and the applications I
needed to do my work. The best name on a leaderboard was not necessarily
the best workable configuration for that afternoon.

\textbf{Third, make session continuity an application-level
responsibility.} Keep prefixes stable where possible, preserve affinity
to reusable cache state and account for auxiliary requests. The
developer can see one continuous conversation while the runtime incurs a
costly new prefill.

\textbf{Fourth, evaluate delegation per verified outcome.} Count
worker-produced code separately from verified application, price the
parent specification, inspection and recovery, and include failed
attempts. The observed delegation arrangements retained useful
open-weight contributions, but in the two studies with a direct arm,
delegation did not reduce measured frontier cost per verified success.

\textbf{Fifth, verify artifacts by behavior when practical.} Cheap
static gates have a purpose, yet a contract stricter than the behavior
the application needs can exclude passing code. Parent acceptance is
another separate filter. The post-hoc test did not replace the frozen
scores; it exposed why a \texttt{malformed} label alone does not mean
the implementation was behaviorally wrong.

\textbf{A follow-up I have not run.} The signature diagnostic points to
one concrete next experiment, which this report does not include. Under
the same fixed contract, a worker whose artifact fails validation would
receive the specific validation error and be allowed one bounded repair
attempt. The study would measure artifact acceptance, genuine verified
worker substitutions, overall task verification, the additional worker
time and inference cost, and frontier cost per verified success with
unsuccessful attempts included. A repair turn might recover some
contract-rejected work, but it also adds parent and worker turns, so it
may not improve verification or reduce cost; only a run with the same
accounting could show which effect dominates.

\textbf{Beyond the rules.} Working with larger cloud-hosted models
sometimes felt different in reasoning and agent behavior, but that was
my impression, not a matched comparison. Those checkpoints, many of them
mixture-of-experts, were used with different machines, runtimes and
clients; I cannot infer a model-size threshold, including one at 70B
parameters, or rank dense and MoE architectures. Larger dense
checkpoints raise a distinct practical resource burden on everyday
developer hardware, while active-parameter counts alone can obscure the
weight-memory burden of an MoE model.

The unfavorable economics should not obscure the value I obtained from
open weights. I could inspect and modify checkpoint choices,
quantization, runtime builds, templates, parsers, cache behavior and
routing rather than wait for a provider's implementation. In the first
study, 26 of 30 delegated trials retained verified worker code; in the
Nemotron study, 11 of 45 were verified worker substitutions under the
stricter contract. Those are bounded but real achievements.
\textbf{Useful local work and lower frontier expenditure proved to be
different questions.}

\subsection{12. Limitations and Evidence
Quality}\label{limitations-and-evidence-quality}

\textbf{Evidence classes are not interchangeable.} A recorded change
shows that something was changed, not what it achieved. A test script's
PASS verdict shows what that script reported, not independent
correctness. Notes written at the time describe what I saw but cannot
replace saved task artifacts. The report follows my evolving workloads
and is not a reproducible controlled comparison across runtimes or
clients.

\textbf{Private-workload boundaries.} Some integration tests and one
observed session used an existing private codebase. They are presented,
where used, only as generic aggregate or qualitative experience; no
source code, prompts, paths or identifiers are disclosed. Their use
remains subject to the author's publication-permission review, and the
entire narrative remains supportable from synthetic cases, October logs
and verified task records if those observations must be removed.

\textbf{Distributed-phase measurements.} I have records of what I built
and decided during the early multi-machine phase but no matched
throughput or latency measurements that would justify numerical
performance claims.

\textbf{AI-assisted task and scorer development.} The author reports
that Claude or another AI assistant contributed to drafting the five
task definitions and hidden tests. The combined harness file was created
about 39 seconds before the first evaluated session and first saved to
version control after the last of the 45 sessions. Task instructions and
visible checks were identical in all 45 saved runs, and recorded score
labels agree with the scorer as later saved. However, no independently
preserved pre-run snapshot of the hidden-test implementation is
available. Changes to hidden scoring logic between file creation and
that later save cannot be ruled out. The precise AI tool and extent of
human review have not been recovered. The study was not preregistered,
and these uncertainties limit claims about benchmark independence. The
second suite was also developed with AI assistance, but its tasks,
hidden checks, worker settings and model revisions were frozen and
hash-recorded before scoring.

\textbf{Author-operated session provenance.} The author confirms that
the 4,165 gateway turns came exclusively from the author's own research
sessions, not other users' sessions. This is a retrospective convenience
sample of the author's work, not a representative production-user
cohort. Personal operation of the sessions does not remove possible
third-party or confidential information embedded in prompts, source
excerpts, and tool results. Only aggregate results are reported.

\textbf{Limited task diversity.} The multi-trial benchmark covered five
selected TypeScript tasks. Additional pilots are one-off fixtures. My
data lack a controlled set of real website-building or broadly
representative software projects. Claims about full-product development
would be unjustified.

\textbf{No front-end design metric.} The verified coding tasks do not
include a validated visual-design, end-user usability, or full-product
experience metric. ``Real coding work'' refers to the agent integration
experience and bounded task-checked implementations, not a broad
assessment of complete application quality.

\textbf{Limited worker diversity in verified tasks.} The initial study
used one Qwen IQ3/llama.cpp worker configuration. The later suite used
Nemotron and Qwen MLX 4-bit configurations on different days. Other
models I configured were not evaluated with comparable hidden checks,
and this does not establish cross-family coding rankings.

\textbf{Fixed parent harness and later-study design.} All delegation
studies used Claude Code with Claude Opus 5.5 as parent. Other harnesses
such as OpenCode were not tested as parents and might specify, inspect
and recover differently. The later studies ran at different times on one
laptop without resetting the parent's prompt cache. Sixteen spend-limit
rejections before inference were replaced contrary to the written
keep-all-sessions rule; keeping those pre-inference rejections as
failures would give direct coding 37/45 and Nemotron 8/45 verified (8
substitutions), rather than the reported replacement-cohort 45/45 and
12/45. One interrupted post-proposal session was retained as a failure.
The Claude Code client version also changed during Nemotron execution.
These circumstances limit causal comparisons.

\textbf{Earlier worker inputs and retrospective sensitivity.} In the
45-trial study, the worker did not receive the parent's separately
provided constraint list; two reconstructed worker prompts were
truncated at 1,800 characters. Both were in failing trials. The later
underscore-relaxation check selected rejected artifacts after the fact,
bypassed one validation gate and did not replay parent review or
application. It is diagnostic, not a revised operational score.

\textbf{Nonrandomized observational data.} Gateway turns and debug
sessions were saved selectively during engineering. Cold and warm cache
cohorts differ in evaluated prefix work and may differ in other aspects.
TTFT differences are descriptive rather than causal.

\textbf{Incomplete hardware and runtime metadata.} Some source files
lack a machine field, exact artifact version, or runtime version.
Machine details I recovered later cannot be reliably matched to
individual saved trials. This prevents hardware-normalized inference
comparisons. Available memory and background application activity were
not logged consistently, so the developer-workstation objective should
not be interpreted as a controlled test of multitasking performance.

\textbf{Price accounting scope.} The cost tables report API
list-price-equivalent frontier cost by token/cache category, not amounts
invoiced. Hardware, engineering time, energy, and depreciation were not
measured. Hence ``cheaper'' refers only to the specified API
list-price-equivalent frontier-cost comparison.

\textbf{Evidence corrections and missing records.} One fee-substitution
attribution was corrected after the run, and both versions were kept.
Some original Nemotron debug details were lost; unrecoverable fields
must remain unknown. The original result files were marked as
non-official evaluation records and should not be recast as a
preregistered official study.

\textbf{No causal evaluation of cache changes.} Engineering rules
describe how to protect exact-prefix reuse, and traffic shows striking
cold/warm differences, but the evidence does not directly quantify the
causal benefit of each patch.

\subsection{13. Related Work}\label{related-work}

SWE-bench {[}1{]} established repository-level issue resolution as a
more realistic coding evaluation than function synthesis, and SWE-Lancer
{[}2{]} connected software tasks to economic value and independent
end-to-end checking. This report extends neither dataset. Instead it
studies operational and economic conditions surrounding smaller,
self-hosted workers, under its own much narrower tasks.

In inference systems, vLLM and PagedAttention {[}3{]} address efficient
KV memory management; SGLang {[}4{]} describes structured LM execution
and RadixAttention-based prefix reuse. This report does not claim either
caching method as original. It documents application-layer
prompt-stability difficulties and operational latency associated with
cache state in long agent sessions. Cost-aware model cascades such as
FrugalGPT {[}5{]} provide conceptual context; a supervised code
delegation workflow is a distinct accounting unit because the parent
still constructs a specification and verifies the artifact.

Recent research closely overlaps with parts of this study. Jarmak
{[}9{]} synthesizes evidence that coding-agent outcomes depend on a
reliability chain extending well beyond the model, while Harness-Bench
{[}11{]} and Fan et al.~{[}12{]} experimentally document harness
sensitivity and tool-interface effects. I do not claim those general
conclusions as new. A non-peer-reviewed single-H100 report by
serbulent-av {[}10{]} evaluates Qwen3-Coder-30B-A3B and Gemma-4-26B-A4B
in agentic coding. Its plan-once and review-loop experiments also
underscore that orchestration choices interact with the worker and test
setup; those pilots did not evaluate the same delegation contract or
frontier-cost accounting used here. Its single-H100 results should not
be conflated with its separate multi-GPU (8$\times$ H200) evaluation.

Peng, Lin, and Lee {[}6{]} examine cloud versus on-premise coding-agent
economics in a nonrandomized longitudinal case study. Their
cache-intensive frontier API configuration was economical per token;
their proposed hybrid routing gateway was evaluated by offline
simulation rather than execution. By contrast, my narrower structured
study executed a frontier-parent/local-worker delegation contract and
accounted for the parent model's frontier cost per task-verified
outcome, though not local hardware cost. Scrouting {[}13{]} reports a
positive but reverse-direction routing design, making clear why my
negative result should not be interpreted as a rejection of cascades
generally. Ari\~{n}o de la Rubia and Pafka {[}7{]} find substantial
run-to-run variability across open-weight agent configurations and
cache-sensitive cost differences across a much larger repeated-run
study; their results underscore the limited precision of my three-trial
cells. Nixon and colleagues {[}8{]} report strong results for local
open-weight agents in a governance-restricted research-data-preparation
setting; their evaluation is restricted to that domain.

Relative to these studies, my contribution is an evidence-qualified
account of \textbf{(1)} model/runtime and coding-client interoperability
across a July-to-October engineering journey, \textbf{(2)}
application/runtime cache and version-management observations from
self-hosted inference, and \textbf{(3)} counted gateway and protocol
failure events, complemented by \textbf{(4)} verified delegation across
two task suites and three open-weight worker configurations from two
model families, with separate accounting for validation, parent review
and frontier spending. The contribution is the combined engineering
account and its bounded studies, rather than a new general-purpose
benchmark.

\subsection{14. Conclusion}\label{conclusion}

The central lesson of this engineering journey was the distance between
\textbf{running an open-weight model} and \textbf{reliably getting
software work done through a coding agent}. I started by trying to share
inference across ordinary machines. I learned to make deployments
repeatable, tool calls executable and sessions less fragile, and finally
to ask whether the code left behind actually passed independent checks.
The same nominal API could hide different model behaviors, runtime
states and interface failures. Neither a successful startup nor a
persuasive explanation was enough.

The structured studies gave a bounded but encouraging answer to the
title's question: \textbf{yes, open-weight workers produced verified
coding contributions}, including 26 retained-code successes in 30
delegated trials in the first study and 11 verified substitutions in 45
Nemotron delegated sessions under a stricter contract. They also exposed
real limits. Under that contract Qwen recorded no end-to-end verified
success; parent review rejected additional proposals; a static
name-preservation rule rejected 19 artifacts that passed the original
hidden checks when applied unchanged offline. In the two studies with a
direct arm, direct frontier execution had the lower measured frontier
cost per verified result. These are properties of the tested
combinations, not final verdicts on the models or on delegation itself.

Despite those difficulties, I finish more optimistic about open-weight
AI than when I began. Over roughly three months of hands-on work, I saw
new capabilities, improvements in runtimes and tooling, and recurring
integration problems become easier to understand or address. Those
observations do not establish a measured rate of industry-wide progress,
and the path was rarely linear. Yet the direction felt increasingly
promising: the distance between a model that runs and a system that does
useful work became more tangible and, in places, more manageable.

That progress rests on a remarkable shared effort: researchers releasing
open weights, engineers improving kernels and serving systems,
maintainers fixing templates and runtime incompatibilities, and
contributors to projects such as llama.cpp, MLX, vLLM and SGLang. Much
of their work is invisible in benchmark tables, but I encountered its
value daily. I am grateful to the people making these systems
accessible, inspectable and improvable.

\textbf{This report records the difficulties of one moment in that
progress, not a permanent boundary. The gap between generating code and
delivering trustworthy work remains substantial; the reason for optimism
is how much thoughtful work is going into closing it.}

\subsection{References}\label{references}

{[}1{]} Jimenez, C. E., et al.~(2024). \emph{SWE-bench: Can Language
Models Resolve Real-world GitHub Issues?} International Conference on
Learning Representations.
\url{https://proceedings.iclr.cc/paper_files/paper/2024/hash/edac78c3e300629acfe6cbe9ca88fb84-Abstract-Conference.html}

{[}2{]} Miserendino, S., Wang, M., Patwardhan, T., and Heidecke, J.
(2025). \emph{SWE-Lancer: Can Frontier LLMs Earn \$1 Million from
Real-World Freelance Software Engineering?} ICML, Proceedings of Machine
Learning Research 267.
\url{https://proceedings.mlr.press/v267/miserendino25a.html}

{[}3{]} Kwon, W., et al.~(2023). \emph{Efficient Memory Management for
Large Language Model Serving with PagedAttention.} SOSP.
\url{https://doi.org/10.1145/3600006.3613165}

{[}4{]} Zheng, L., et al.~(2024). \emph{SGLang: Efficient Execution of
Structured Language Model Programs.} NeurIPS 37.
\url{https://proceedings.neurips.cc/paper_files/paper/2024/hash/724be4472168f31ba1c9ac630f15dec8-Abstract-Conference.html}

{[}5{]} Chen, L., Zaharia, M., and Zou, J. (2024). \emph{FrugalGPT: How
to Use Large Language Models While Reducing Cost and Improving
Performance.} Transactions on Machine Learning Research.
\url{https://openreview.net/forum?id=cSimKw5p6R} (also arXiv:2305.05176)

{[}6{]} Peng, S.-W., Lin, Y.-H., and Lee, Y.-P. (2026). \emph{Inference
Economics of Enterprise Coding Agents: A Case Study of Cloud
vs.~On-Premise LLMs.} arXiv:2607.13080.
\url{https://arxiv.org/abs/2607.13080}

{[}7{]} Ari\~{n}o de la Rubia, E., and Pafka, S. (2026). \emph{Identical
Runs, Different Results: Benchmarking AI Coding Agents on Open-Weight
Models.} arXiv:2609.33812. \url{https://arxiv.org/abs/2609.33812}

{[}8{]} Nixon, M., et al.~(2026). \emph{Agentic coding without the
cloud: evaluating open-weight large language models on longitudinal data
preparation tasks.} arXiv:2607.21482v2.
\url{https://arxiv.org/abs/2607.21482v2}

{[}9{]} Jarmak, S. (2026). \emph{Engineering Reliable Coding Agents:
Evaluating and Operating the System Around the Model.}
arXiv:2608.13867v1. \url{https://arxiv.org/abs/2608.13867v1}

{[}10{]} serbulent-av (2026). \emph{Self-Hostable Open-Weight Coding
Models on a Single H100.} GitHub technical report,
\texttt{benchmarks/local-llm-coding-benchmark.md}, commit
\texttt{8098743985f3} (not peer-reviewed).
\url{https://github.com/serbulent-av/agentic-coding/blob/8098743985f3/benchmarks/local-llm-coding-benchmark.md}

{[}11{]} Yao, Y., et al.~(2026). \emph{Harness-Bench: Evaluating Agent
Harnesses across Models and Realistic Workflows.} arXiv:2605.27922v2 (15
October 2026). \url{https://arxiv.org/abs/2605.27922v2}

{[}12{]} Fan, R.-Z., et al.~(2026). \emph{An Empirical Study of Harness
Design for Coding Agents.} arXiv:2609.20804v1.
\url{https://arxiv.org/abs/2609.20804v1}

{[}13{]} Bhola, I., Krishnan, A., and NS, M. (2026). \emph{Scrouting:
Cost-Aware Routing of Coding Agents by Scouting the Repository First.}
arXiv:2608.04804v1. \url{https://arxiv.org/abs/2608.04804v1}

\subsection{Statements and data
availability}\label{statements-and-data-availability}

\textbf{Data availability.} The underlying research records are
privately preserved and are not part of this publication. No raw gateway
logs, coding-agent or Cursor transcripts, prompts, model responses,
source code (including harness code), deployment configurations, debug
archives, evidence manifests, benchmark result files, task fixtures,
hidden tests, scorers, or other raw research artifacts are publicly
released or offered on request. The paper reports aggregated results and
describes the methodology where possible, but independent reproduction
and audit are limited by the restricted availability of the underlying
records. A sanitized subset of the synthetic fixtures, evaluation checks
and aggregate results may be made available following privacy,
intellectual-property and publication-permission review. No public
release is currently authorized.

\textbf{Use of AI-assisted tools.} AI-assisted tools were used in
developing portions of the experimental harness, auditing evidence,
checking reported statistics, and preparing and reviewing this
manuscript. The author directed the research, evaluated the evidence,
and takes responsibility for the reported findings and conclusions.

\textbf{Competing interests.} The author is developing self-hosted AI
infrastructure with potential commercial applications. The experimental
system examined in this report is related to that work.

\textbf{Computing support.} Computing resources were supported in part
by Google Cloud Free Trial credits.

\end{document}